\documentclass[a4paper,fleqn,usenatbib]{mnras}

\usepackage{newtxtext,newtxmath}
\usepackage[T1]{fontenc}
\usepackage{ae,aecompl}

\usepackage{graphicx}	% Including figure files
\usepackage{amsmath}	% Advanced maths commands
\def\spose#1{\hbox to 0pt{#1\hss}}
\def\lta{\mathrel{\spose{\lower 3pt\hbox{$\mathchar"218$}}
     \raise 2.0pt\hbox{$\mathchar"13C$}}}
\def\gta{\mathrel{\spose{\lower 3pt\hbox{$\mathchar"218$}}
     \raise 2.0pt\hbox{$\mathchar"13E$}}}

\def\p0{\phantom{0}}

\title[Chromospheric Activity in 55~Cancri~II.]
{Chromospheric Activity in 55~Cancri: \\
II.~Theoretical Wave Studies versus Observations}

\author[Cuntz et al.]{
Manfred Cuntz$^{1}$\thanks{E-mail: cuntz@uta.edu (MC)}, Klaus-Peter Schr\"oder$^{2,3}$, Diaa E. Fawzy$^{4}$, and Andrew R. Ridden-Harper$^{5}$
\\
$^{1}$Department of Physics, University of Texas at Arlington, Arlington, TX 76019, USA \\
$^{2}$Departamento de Astronom{\'\i}a, Universidad de Guanajuato, A. P. 144, 36000 Guanajuato, GTO, Mexico \\
$^{3}$Sterrewacht Leiden, Nils Bohrweg 2, NL-2333 CA Leiden, The Netherlands \\
$^{4}$Faculty of Engineering, Izmir University of Economics, 35330 Izmir, Turkey \\
$^{5}$Department of Astronomy and Carl Sagan Institute, Cornell University, Ithaca, NY 14853, USA
}

\date{Accepted XXX. Received YYY; in original form ZZZ}

\pubyear{2020}

\begin{document}
\label{firstpage}
\pagerange{\pageref{firstpage}--\pageref{lastpage}}
\maketitle

  \begin{abstract}
In this study, we consider chromospheric heating models for 55~Cancri in
conjunction with observations.  The theoretical models, previously discussed
in Paper~I, are self-consistent, nonlinear and time-dependent {\it ab-initio}
computations encompassing the generation, propagation, and dissipation of waves.
Our focus is the consideration of both acoustic waves and longitudinal flux tube
waves amounting to two-component chromosphere models.  55~Cancri,
a K-type orange dwarf, is a star of low activity, as expected by its age, which
also implies a relatively small magnetic filling factor.  The Ca~II~K fluxes are computed
(multi-ray treatment) assuming partial redistribution and time-dependent ionization.
The theoretical Ca~II H+K fluxes are subsequently
compared with observations.  It is
found that for stages of lowest chromospheric activity the observed Ca~II fluxes are
akin, though not identical, to those obtained by acoustic heating, but agreement can be
obtained if low levels of magnetic heating --- consistent with the assumed photospheric
magnetic filling factor --- are considered as an additional component; this idea is in
alignment with previous proposals conveyed in the literature.
  \end{abstract}


%%%%%%%%%%%%%%%%%%%%%%%%%%%%%%%%%%%%%%%%%%%%%%%%%%

%%%%%%%%%%%%%%%%% BODY OF PAPER %%%%%%%%%%%%%%%%%%

\section{Introduction}

This study is a continuation of our previous work aimed at describing
chromospheric heating in late-type stars by considering both theoretical
wave models and observations.  A large body of previous work, see, e.g.,
\cite{sch00} and references therein, indicates that the atmospheres of highly
active stars are dominantly heated by magnetic processes, including magnetic waves
\citep[e.g.,][]{sho20},
whereas for the atmospheric heating of low-activity stars non-magnetic processes
are expected to play a more prominent role \citep[e.g.,][]{sch95,buc98,cun99},
even though magnetic phenomena.  When stars age, the relative importance of
atmospheric magnetic processes tends to subside, a process closely related
to the evolution of angular momentum \citep[e.g.,][]{kep95,cha97,wol97,mat15,mit18}.

Here we focus on 55~Cancri (55~Cnc, $\rho^1$~Cnc), a G8~V star \citep{gon98}
of advanced age, also identified as an orange dwarf.  Orange dwarfs are of
particular interest to a large range of astrophysical studies as, for example,
they constitute transitory objects regarding various kinds of atmospheric
heating processes.  55~Cnc is an example of a slow rotating star as
its slow rotation is closely connected to its advanced age;
see, e.g., \cite{sku72} and subsequent work, including \cite{bar03,bar07},
for background information.  Following
\cite{faw21}, see Paper~I, this information is relevant for constraining the
stellar photospheric and chromospheric magnetic filling factor
(MFF\footnote{See Table~1 for a summary of the acronyms.}), in part based
on empirically deduced statistical relationships \citep[e.g.,][and references
therein]{cun99}.  Additionally, information on the magnetic and nonmagnetic
heating parameters has been provided as well.

Orange dwarfs are of particular interest to both astrophysics and astrobiology
in consideration of various features, including those commonly considered favorable
in support of exolife \citep[e.g.,][]{cun16,lin18,lin19,dvo20}.  Those include the relative
frequency of those stars (if compared to stars akin to the Sun), the relatively
large size of their habitable zones (HZs) (if compared to M dwarfs), and their long main-sequence
life times (i.e., 15~Gyr to 30~Gyr, compared to about 10~Gyr for solar-like stars;
see, e.g., \citeauthor{pol98} \citeyear{pol98}).  Recent work on the significance
of late G and K dwarfs for the possibility of supporting exolife has been given by,
e.g., \cite{arn19} and \cite{sch20}.

Previous studies about 55~Cnc also entertain the possible existence of Earth-type planets
as part of the 55~Cnc system, especially in the region between 0.8~au and 5.7~au
from the star.  This spatial domain allows terrestrial planets to exhibit long-term
orbital stability \citep[e.g.,][]{ray08,sat19}.  Another pivotal motivation for the
study of outer atmospheric heating and emission in 55~Cnc and similar stars is set
by the emergent field of space weather simulations.  This kind of work helps to
augment our understanding of stellar and planetary environments, besides fostering
basic astrophysical research \citep[e.g.,][]{lam12,air20}.

Our paper is structured as follows: In Section~2, we convey our theoretical approach,
including comments on the time-dependent wave computations.  In Section~3, we
discuss observational aspects, including the calibration of the observed spectrum.
Results and discussions are given in Section~4, including a discourse on the comparison
between the theoretical models and observations.  In Section~5, we report our summary
and conclusions.

%++++++++++++++++++++++++++++++++++++++++++++++++++++++++++++++++++++++++++++++++
%++++++++++++++++++++++++++++++++++++++++++++++++++++++++++++++++++++++++++++++++

\section{Theoretical Approach}

%++++++++++++++++++++++++++++++++++++++++++++++++++++++++++++++++++++++++++++++++

\subsection{Stellar Parameters}

We pursue atmospheric studies of 55~Cnc, a G8~V star \citep{gon98} that has
a mass of about 0.91~$M_\odot$ \citep{bra11}.  Its effective temperature has been
determined as $5165{\pm}46$~K \citep{lig16} and $5172{\pm}18$~K \citep{yee17}.
Furthermore, 55~Cnc is notably older than the Sun with previous estimates given as
8.6~Gyr \citep{mam08} and 10.2~Gyr \citep{bra11}; see also \cite{bou18} for discussion.
Previous work about 55~Cnc's rotation period indicates values of 42.2~d \citep{hen00}
and 38.8~d \citep{bou18}; clearly, these values of slow rotation are closely connected to
55~Cnc's age, see, e.g., \cite{sku72} and subsequent work.
As pointed out in Paper~I, the stellar rotation period also allows default estimates
of the stellar photospheric magnetic filling factor (see Figure~1).

In the context of chromospheric heating, increased amounts of magnetic heating (in a
statistical sense) are strongly correlated with increased outer atmospheric emission
\citep[e.g.,][]{lin83}, and related studies.  Moreover, as discussed by \cite{cun98,cun99}
and references therein, it is also possible to link $B_{0}f_{0}$ to both the stellar rotation period
$P_{\rm rot}$ \citep[e.g.,][]{mar89,mon93,saa96a,rei09} as well as to the emergent chromospheric
emission flux \citep[e.g.,][]{saa87,sch89,mon93,saa96b,jor97}.  Related theoretical studies
for sets of main-sequence stars, including objects akin to 55~Cnc, have been given by
\cite{faw98,faw02b}.

An important aspect concerning the magnitude of stellar activity, as given by the Ca~II emission,
pertains to the Rossby number (i.e., the ratio between the stellar rotational period and
the local convective turnover time, $\tau_{\rm c}$).  The Rossby number is related to the dynamo process;
it also exhibits an observational correlation with the stellar activity.  Based on work by,
e.g., \cite{mon01}, \cite{cra11}, and \cite{cas14}, $\tau_{\rm c}$ for 55~Cnc is given as
about 21.5~d; thus, the Rossby number for 55~Cnc is identified as ${\rm Ro} \simeq 1.9$,
consistent with a star of low chromospheric activity; see Appendix~A for details.  \cite{noy84}
previously identified correlations between $P_{\rm rot}$ and the chromospheric basal flux emission,
showing well-pronounced trends albeit considerable scatter with respect to the observational data.

Empirically, the decrease of activity in main-sequence stars does not necessarily concur with their absolute age,
but rather with their relative main-sequence age as the magnetic braking time-scale increases downward
on the main-sequence akin to their main-sequence evolutionary life time \citep{rei12,sch13}.  But 55 Cnc
is more evolved than the Sun even on a relative time-scale, in consideration of that 55 Cnc has
passed about 2/3 of its time on the main-sequence (which is about 15 Gyr), contrary to the Sun with
an age of about mid main-sequence.  Hence, 55 Cnc is indeed a low-activity star while being quite evolved.

%++++++++++++++++++++++++++++++++++++++++++++++++++++++++++++++++++++++++++++++++

\subsection{Time-dependent Wave Calculations}

In Paper~I, we pursued theoretical wave calculations based on
acoustic waves (ACWs) and longitudinal flux tube waves (LTWs).  Previously
developed codes allowed us addressing the different steps of the magneto-acoustic
heating picture, comprised of convective energy generation \citep[e.g.,][]{mus94,mus95,ulm96},
the propagation and dissipation of waves through different layers of the stellar atmosphere
\citep[e.g.,][]{buc98,faw02a}, and the emergence of radiative energy output \citep[e.g.,][]{cun99,faw02a}.
Both acoustic and magnetic waves are followed starting from the stellar convective zone to the point
of shock formation and beyond.  As discussed in Paper~I, the initial wave energy fluxes for ACWs
and LTWs have been chosen as $F_{\rm ACW} = 3.3 \times 10^7$ erg~cm$^{-2}$~s$^{-1}$ and
$F_{\rm LTW} = 1.7 \times 10^8$ erg~cm$^{-2}$~s$^{-1}$, respectively.  We examined models
based on monochromatic waves and frequency spectra.  In case of monochromatic waves,
we chose $P=60$~s as representative period.  This value is motivated by the shapes of the
ACW and LTW frequency spectra as in both spectra the wave energy flux reaches its maximum
close to that value (see Paper~I).

The photospheric MFF is set by an empirical relationship based on observed values for the stellar
rotation period, $P_{\rm rot}$; see \cite{cun99}.  Assuming 42.2~d \citep{hen00} and 38.8~d \citep{bou18}
amount to $f_0=0.3$\% and $1.4$\%, respectively.  The adopted MFF also informs the shape
of the magnetic flux tubes as considered for our studies (see Fig.~1).  The chromospheric
Ca~II fluxes are computed by considering non-local thermodynamic equilibrium (NLTE) and
partial redistribution (PRD).  In addition, effects originating from time-dependent ionization (TDI)
have been taken into account as well; see \cite{ram03} and \cite{faw12} for details.  The
self-consistent treatment of TDI (especially for hydrogen) greatly impacts the atmospheric
temperatures and electron densities (especially behind the shocks); it also affects the
emergent Ca II fluxes.  See Fig.~2 for an example of a snapshot series pertaining to
the dynamics associated with LTW spectral waves, including information on the Ca~II/Ca
and Mg~II/Mg ionization degrees.

For the adequate representation of the multi-level atomic model for Ca~II, we rely on previous
results by \cite{faw15}.  In order to model the total radiative energy losses, adequate
scaling factors are applied to both the Ca~II~K and Mg~II~{\it k} lines.  Radiative energy
fluxes for our two-component chromosphere models, based on both ACWs and LTWs, are
computed through implementing multi-ray treatments.  For selected sets of models, increased
initial wave energy fluxes are used to mimic the impact of torsional flux tube waves (TTWs)
regarding the wave energy budget as expected through the occurrence of mode coupling
\citep[e.g.,][]{hol82,kud99,bog03,has03}.

%++++++++++++++++++++++++++++++++++++++++++++++++++++++++++++++++++++++++++++++++

%  Table 1
%%%%%%%%%%%%%%%%%%%%%%%%%%%%%%%%%%%%%%%%%%%
%
\begin{table}
	\centering
	\caption{Summary of Acronyms}
\begin{tabular}{ll}
\noalign{\smallskip}
\hline
\noalign{\smallskip}
Acronym  &  Definition  \\
\noalign{\smallskip}
\hline
\noalign{\smallskip}
ACW     &  Acoustic Wave                                \\
HZ      &  Habitable Zone                               \\
LTE     &  Local Thermodynamic Equilibrium              \\
LTW     &  Longitudinal Flux Tube Wave                  \\
MFF     &  Magnetic Filling Factor ($f_0$)              \\
NLTE    &  Non-Local Thermodynamic Equilibrium          \\
NTDI    &  Non-Time-Dependent Ionization                \\
PRD     &  Partial Redistribution                       \\
TDI     &  Time-Dependent Ionization                    \\
TTW     &  Torsional Flux Tube Wave                     \\
\noalign{\smallskip}
\hline
\end{tabular}
\end{table}
%%%%%%%%%%%%%%%%%%%%%%%%%%%%%%%%%%%%%%%%%%%

%  Figure 1
%%%%%%%%%%%%%%%%%%%%%%%%%%%%%%%%%%%%%%%%%%%
%
\begin{figure}
\centering
  \includegraphics[width=1.0\linewidth]{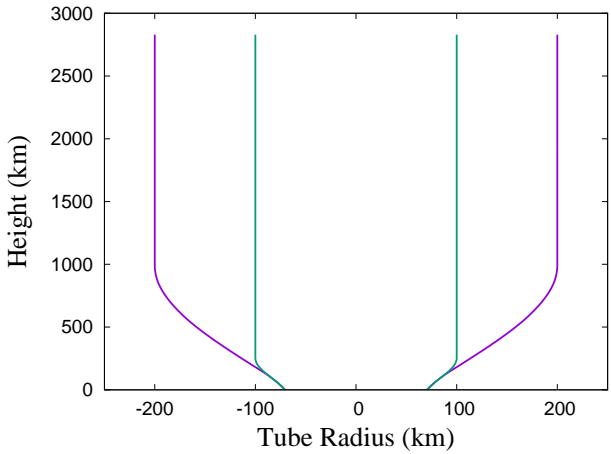}
  \caption{Initial magnetic flux tube models for 55~Cnc.  The photospheric
magnetic filling factor (main models) is given as $f_0$ = 0.3\% (purple).
We also discuss some models with $f_0$ = 1.4\% for comparison (green).
The tube opening radii are given as 100~km and 200~km, respectively;
see Paper~I for further information.}
\end{figure}
%%%%%%%%%%%%%%%%%%%%%%%%%%%%%%%%%%%%%%%%%%%

%  Table 2
%%%%%%%%%%%%%%%%%%%%%%%%%%%%%%%%%%%%%%%%%%%
%
\begin{table}
	\centering
	\caption{Results for the LTW TDI Mechanical Energy Flux}
\begin{tabular}{rcccc}
\noalign{\smallskip}
\hline
\noalign{\smallskip}
Height   & \multicolumn{2}{c}{LTW} & \multicolumn{2}{c}{2 $\times$ LTW} \\
\noalign{\smallskip}
\hline
\noalign{\smallskip}
  ...    & $f_0=0.3$\% & $1.4$\% & $0.3$\% & $1.4$\% \\
(km)     & (cgs)       & (cgs)   & (cgs)   & (cgs)   \\
\noalign{\smallskip}
\hline
\noalign{\smallskip}
\multicolumn{5}{c}{Monochromatic Wave Models}  \\
\noalign{\smallskip}
\hline
\noalign{\smallskip}
0      &  8.23  &  8.23  &  8.53  &  8.53  \\
250    &  7.60  &  7.64  &  7.85  &  7.89  \\
500    &  7.24  &  7.52  &  7.45  &  7.72  \\
750    &  6.81  &  7.10  &  6.96  &  7.26  \\
1000   &  6.22  &  6.49  &  6.36  &  6.63  \\
1250   &  5.57  &  5.80  &  5.70  &  5.94  \\
1500   &  4.92  &  5.12  &  5.04  &  5.25  \\
1750   &  4.35  &  4.49  &  4.44  &  4.61  \\
2000   &  3.84  &  3.94  &  3.92  &  4.05  \\
2250   &  3.38  &  3.49  &  3.46  &  3.59  \\
2500   &  2.96  &  3.09  &  3.05  &  3.23  \\
\noalign{\smallskip}
\hline
\noalign{\smallskip}
\multicolumn{5}{c}{Spectral Wave Models}  \\
\noalign{\smallskip}
\hline
\noalign{\smallskip}
0      &  8.23  &  8.23  &  8.53  &  8.53  \\
250    &  7.58  &  7.56  &  7.84  &  7.83  \\
500    &  7.18  &  7.47  &  7.40  &  7.67  \\
750    &  6.78  &  7.16  &  7.02  &  7.35  \\
1000   &  6.34  &  6.68  &  6.56  &  6.87  \\
1250   &  5.84  &  6.16  &  6.03  &  6.33  \\
1500   &  5.26  &  5.57  &  5.40  &  5.77  \\
1750   &  4.77  &  5.03  &  4.81  &  5.18  \\
2000   &  4.28  &  4.52  &  4.32  &  4.64  \\
2250   &  3.84  &  4.04  &  3.87  &  4.12  \\
2500   &  3.41  &  3.55  &  3.47  &  3.58  \\
\noalign{\smallskip}
\hline
\noalign{\smallskip}
\multicolumn{5}{p{0.57\columnwidth}}{
Note:
Results are given in logarithmic units.
}
\end{tabular}
\end{table}

%%%%%%%%%%%%%%%%%%%%%%%%%%%%%%%%%%%%%%%%%%%

%++++++++++++++++++++++++++++++++++++++++++++++++++++++++++++++++++++++++++++++++
%++++++++++++++++++++++++++++++++++++++++++++++++++++++++++++++++++++++++++++++++

\section{Observational Aspects}

\subsection{Background Information}

A major factor in the comparison of the theoretical Ca~II~K line
profile and the total Ca~II~K emission with observations,
besides considering the instrumental profile, 
is given by the photon absorption and reemission processes.
The latter occur between the bottom of chromosphere, where the emission
is produced, and the top, where the observed line profile emerges.
In the absence of considerable collisions ($C \ll A$ for the
Ca II H+K transition), owing to the relatively low chromospheric
densities, absorbed photons are almost always reemitted.
Hence, we can expect not to lose any of the produced emission,
and even though the chromosphere is optically thick, it is nonetheless
``effectively optically thin''.  Thus, the computed emission measure
should directly reflect the observed value.

However, the many absorption and reemission processes in the
Ca~II~H and K lines, along the line of sight in consideration
of the chromospheric mass column density, amount to the well-known
Wilson-Bappu effect \citep{wil57}, which in the example studied here
operates in a single chromosphere.  As previously described by \cite{ayr75},
during that process the photons migrate toward the wings of the line
profile, where their escape probability is larger as the
optical depth there is smaller.  Consequently, a line broadening
effect occurs, caused by the high optical depth at the centers of the
Ca~II~H and K lines that depends on the mass column density of
the particular chromosphere.  Consequently, the widening effect
is larger for stars of lower gravity, notably for giant stars.
More luminous giants have a lower gravity, a larger chromospheric
scale height, and, consequently, exhibit broader Ca~II~H and K
emission lines.

The Wilson-Bappu effect can usually only be observed by comparing
stars of different surface gravity.  Here, however, we have the unique
opportunity of discerning this widening process for a given chromosphere,
as we are able to compare the computed line profile produced at the
bottom with the one at the top, noting that the latter given as the
observed emergent line profile.

%++++++++++++++++++++++++++++++++++++++++++++++++++++++++++++++++++++++++++++++++

\subsection{Comparing Initial and Emerging Emission Line Profiles:
  The Wilson-Bappu Effect at Work}

The main focus of this study is to compare the Ca~II~K emission line
profile, generated due to wave heating (see Paper~I) and mostly formed
at the bottom of the chromosphere near the temperature minimum,
with observations.  The observed line spectrum (resolving power: $R$=115.000)
was obtained in 2013 \citep{rid16} at HARPS-N\footnote{The HARPS-N observations
were originally taken in TNG Observing programme CAT13B$_33$ (PI: F. Rodler),
and were used by \cite{lop14} to investigate the Rossiter McLaughlin effect.}.
The associated instrumental line profile has a width of only 0.035~{\AA},
which is more than an order of magnitude narrower than the intrinsic
line width.

The observed emission line profile is, in fact, noticeably wider than that
of the theoretically expected emission, i.e., $W_\circ = 0.50$~{\AA} compared
to, typically, $W_\circ = 0.25$~{\AA} at half peak (based on the description
by O. C. Wilson).  At the base of the emission, the difference is 0.70~{\AA},
as of the observed profile, compared to 0.50~{\AA} for the computed
emission profiles; see Fig.~3 for a direct comparison.  Surely, the
observed profile is also embedded in the relatively wide photospheric
core of that line.  Its residual flux is defined by the temperature of
the stellar atmosphere at and below the temperature minimum region.

Hence, this difference in the emission line width being much too large
to be attributable to instrumental effects, it is apparent that the
line profile is mostly due to the Wilson-Bappu effect operating along
the line of sight.  In particular, the many absorption and reemission
processes each emerging photon undergoes favors the migration into
the wings of the line profile function, resulting in the broadening
of the line (see Sect.~3.1)  The explanation of this process,
as forwarded by \cite{ayr75}, suggests a distinct relationship
between the mass column density $N$ and gravity $g_\star$
($N \propto 1/\sqrt{g_\star}$).  Thus, it yields a dependence
of the line width $W_\circ$ to the star's gravity given as
$W_\circ \propto {g_\star}^{-0.25}$.  A comparison of giant stars
of known physical parameters indicates that this assessment is
largely consistent with observation; see, e.g., \cite{ayr75},
\cite{lut82}, and work in progress by one of us (K.-P.~S.).

In this paper, however, we are able to demonstrate that this effect
also operates in a distinct chromosphere, as we can showcase the
comparison of the observed chromospheric emission line profile with
the one obtained by theoretical modelling.  This unique opportunity
allows us, from a different point of view, to highlight density
broadening as the best explanation for the Wilson-Bappu effect
--- considering that previously an alternative explanation for the
line width, based on Doppler broadening by turbulence, has been
forwarded; see, e.g., \cite{rei73} as well as the extensive
discussion by \cite{lut82}.  So far, this alternate explanation
of the emission line broadening has never been fully ruled out.

In fact, the Wilson-Bappu effect, if operating as discussed, is also
producing the right magnitude of line broadening.  To demonstrate this,
we may simply compare a hypothetical star of twice the chromospheric
column density as 55~Cnc, by having only a quarter of its gravity, 
but otherwise having the same effective temperature and metallicity.
Hence, its Ca~II line emission width $W_\circ$ is increased by a factor of
$g_\star^{-0.25} \propto \sqrt{2}$.  Considering an instrumental profile width of
35~m{\AA}, the true value of 55 Cnc's emission line width $W_\circ$ is expected to be close
to 465~m{\AA}, or by 215~m{\AA} larger than the theoretical emission line width at the
bottom of 55 Cnc's chromosphere, whereas $W_\circ$ of the hypothetical star would
amount to 660~m{\AA}.  In other words, adding the same chromospheric column
density once again in this application, the Wilson-Bappu effect entails an additional
line broadening by the same amount (i.e., about 200~m{\AA}) as the one required
to reconcile the theoretically predicted emission line width with the
observed one considering the radiative transfer regarding 55~Cnc's
chromosphere (see Fig. 3). 

%++++++++++++++++++++++++++++++++++++++++++++++++++++++++++++++++++++++++++++++++

%  Figure 2
%%%%%%%%%%%%%%%%%%%%%%%%%%%%%%%%%%%%%%%%%%%
%
\begin{figure*}
\centering
  \includegraphics[width=0.40\linewidth]{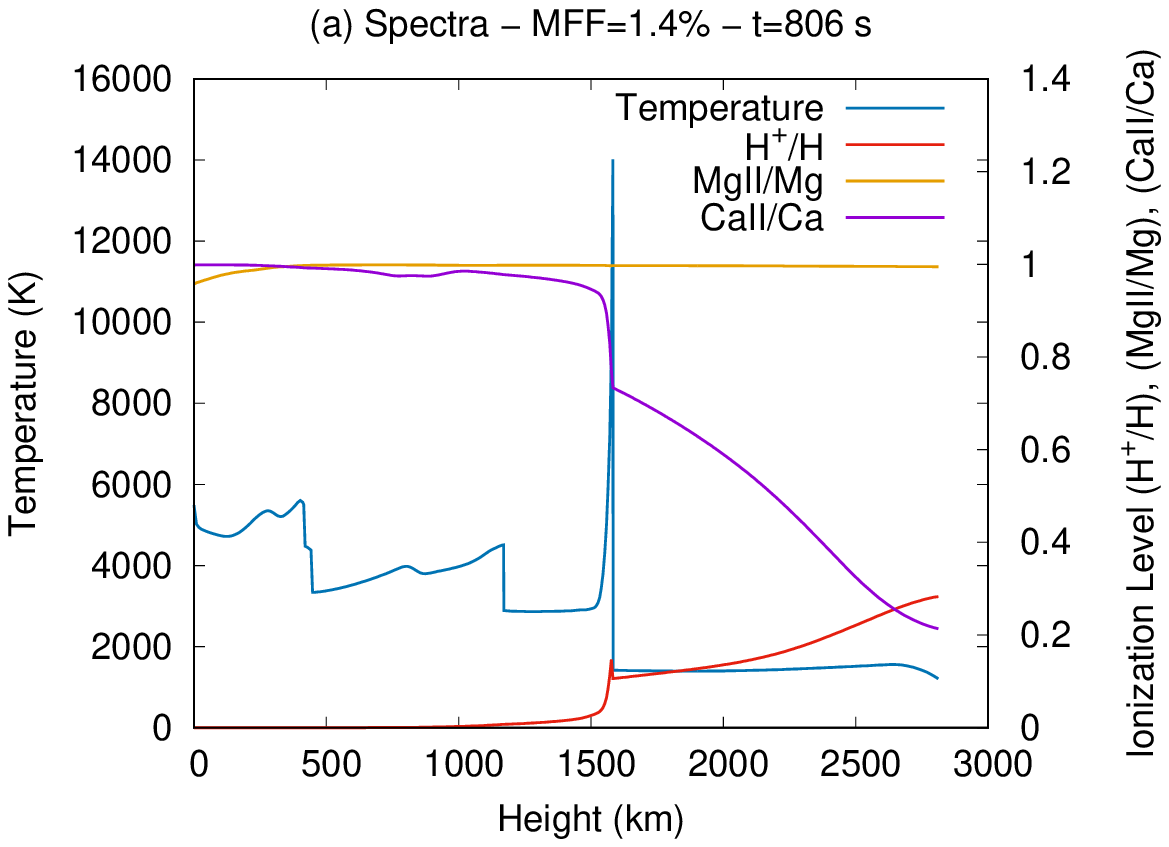}
  \includegraphics[width=0.40\linewidth]{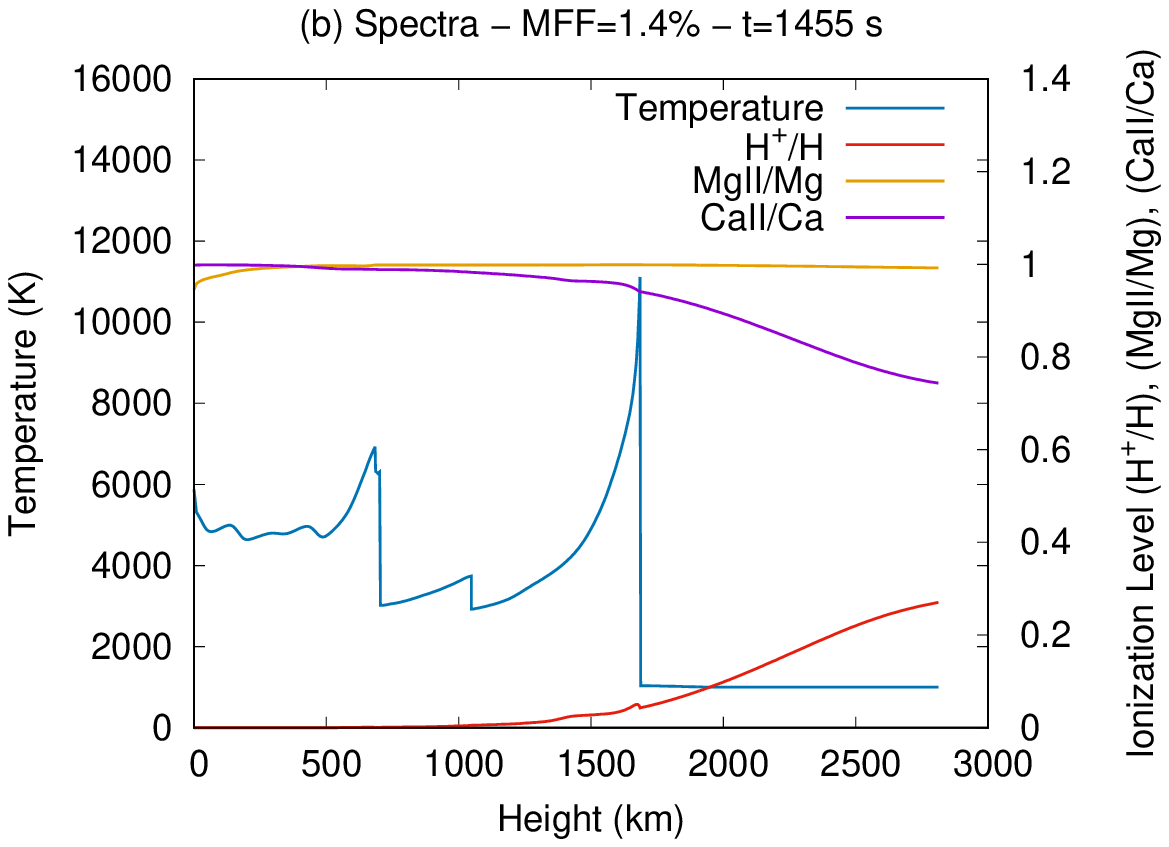} \\
  \includegraphics[width=0.40\linewidth]{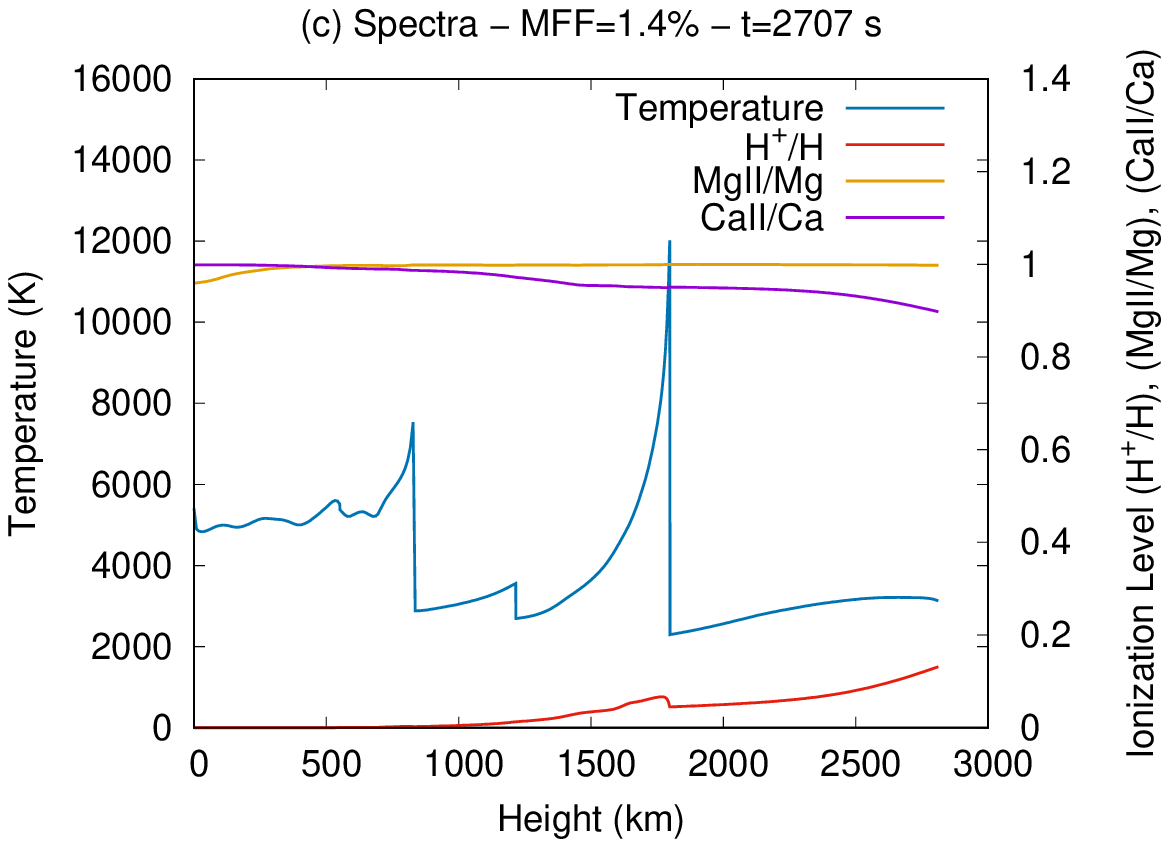}
  \includegraphics[width=0.40\linewidth]{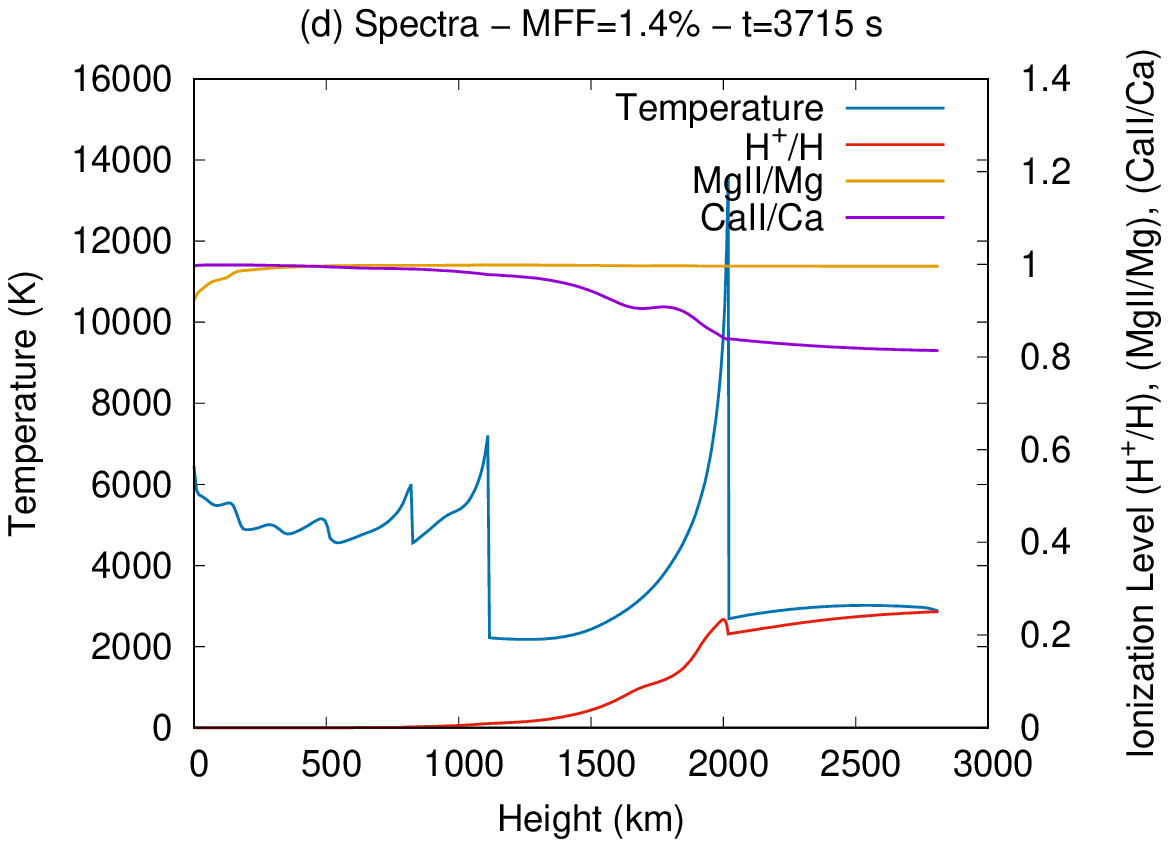}
  \caption{Snapshot series for a LTW TDI model based on spectral waves and a MFF of $f_0=1.4$\%.}
\end{figure*}
%%%%%%%%%%%%%%%%%%%%%%%%%%%%%%%%%%%%%%%%%%%

%  Figure 3
%%%%%%%%%%%%%%%%%%%%%%%%%%%%%%%%%%%%%%%%%%%
%
\begin{figure*}
\centering
  \includegraphics[width=0.60\linewidth]{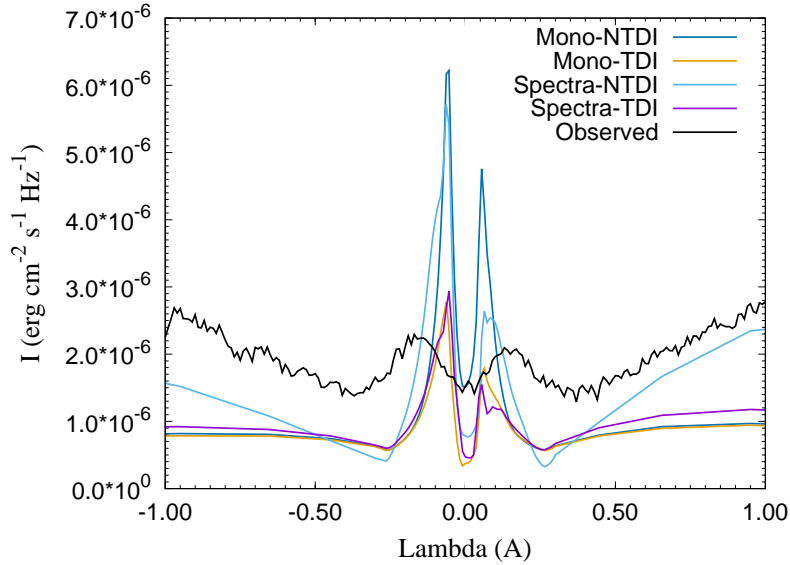}
  \caption{Examples of Ca~II~K profiles for LTWs based on $f_0 = 0.3${\%}
in comparison with observations.}
\end{figure*}
%%%%%%%%%%%%%%%%%%%%%%%%%%%%%%%%%%%%%%%%%%%

\subsection{On the Flux Scale of the Observed Spectrum}

In order to calibrate the scale of the normalized spectral flux of the 
observed Ca~II~K line profile with respect to the spectral surface flux
in absolute terms, we used the best-matching spectrum based on the PHOENIX
atmospheric model\footnote{Sets of synthetic models are publicly
available by the PHOENIX model library at the University of G\"ottingen, see
\cite{hus13}.} based on the parameters for 55 Cnc given as 
$T_{\rm eff}=5200$~K, $\log{g_\star} = 4.0$ (cgs), and [Fe/H]=0.5, which are
consistent with the observed values.  The spectrum as obtained
fits well the relatively wide photospheric Ca II K profile, contrary to, e.g.,
the model based on $T_{\rm eff} = 5100$~K previously used in the literature.

Note that the latter value results in a much wider photospheric Ca~II~K line
profile.  This difference is relevant  for the calibration process as that
cooler model already implies a  20\% lower near-UV spectral flux.  Since the
photospheric wings are a very sensitive indicator of the stellar effective
temperature, we therefore opt in favor of the former value.  In fact,
$T_{\rm eff}$ seems to lie somewhere between  5100~K and 5200~K.
Additionally, the limited resolution of the model library regarding
[Fe/H] and $\log{g_\star}$ made us choose a slightly higher value for [Fe/H]
and a slightly lower value for $\log{g_\star}$ than observed.  Ultimately, effects
on the synthetic spectrum based on these small parameter deviations largely
cancel each other out.

In particular, we hinged the spectral surface flux scale calibration on 
line-free points in the Ca~II~K photospheric profile, from about 2~{\AA}
away of the very core outwards.  Of particular use here is the region
around $-2.5$~{\AA}, which is sufficiently far away from the somewhat
mismatched innermost photospheric line core.  Note that this process
is based on the standard 1-D, NLTE, and photosphere-only models.  
They do not account for the stellar chromosphere entailing that
the temperature minimum is artificially low.  On the other hand,
only the innermost photospheric Ca~II~H+K line cores are formed
at those outermost photospheric layers.  Therefore, the rest of
the PHOENIX Ca~II~K line profile matches the observed spectrum very well.   

Matching the normalized observed flux scale to the above-mentioned PHOENIX 
spectrum including its spectral surface flux scale results in a factor of
$1.75 \times 10^{-5}$; this factor is needed for multiplying the observed
normalized flux to arrive at physical flux units.  It also considers
the different units used by the PHOENIX spectra (i.e., spectral flux
per wavelength in cm), while we use  a spectral flux per frequency in Hz.

Figure~3 conveys the obtained direct and quantitative comparison of the 
observed and computed chromospheric emission on the spectral surface flux scale. 
We estimate the residual uncertainty of the scaling to be on the order of 20\%
based on the expected impact of a small (by 100~K) mismatch of $T_{\rm eff}$.
Moreover, the observed emission rises from the deep photospheric line core,
rather than from flat zero, and it is widened by the above described Wilson-Bappu
effect, i.e., the migration of the photons into the wings of the line profile
function by line saturation effects due to the relatively large Ca~II mass
column density.  Nonetheless, the emission measures --- to be visualized 
as the area of the emission line profile --- are in good agreement.

%++++++++++++++++++++++++++++++++++++++++++++++++++++++++++++++++++++++++++++++++
%++++++++++++++++++++++++++++++++++++++++++++++++++++++++++++++++++++++++++++++++

\section{Results and Discussion}

\subsection{Time-dependent Two-Component Heating Models}

Time-dependent two-component heating models, based on ACWs and LTWs, have been
obtained in Paper~I.  These models are used to calculate the emergent Ca~II H+K fluxes
in correspondence to the various theoretical models.  The most relevant results obtained
in Paper~I include that both ACWs and LTWs form shocks in the upper photosphere /
lower chromosphere.  Regarding monochromatic waves, the height of shock formation
for LTWs (with $f_0 = 0.3$\%) was identified as about 30 km lower than in ACWs due to the
higher wave energy flux. Generally, higher shock strengths are attained for spectral waves
compared to monochromatic waves.  Notable differences between the models occur at
larger heights, mostly due the dilution of the wave energy flux.

In the present study, we revisited this aspect by comparing wave models of different
MFFs and initials wave energy fluxes, given as $F_{\rm LTW}$ and $2 \times F_{\rm LTW}$;
we also assumed TDI (see Table~2).  We found that the wave energy fluxes decreased by
many orders of magnitude as a function of height, as expected.  For a fixed height,
a larger MFF resulted in a higher value for the wave energy flux, a consequence of
the different magnitude of wave energy flux dilution (see Fig.~1).  For example, at
heights of 750~km, the difference between $f_0$ = 0.3\% and 1.4\% resulted in a
factor between 2.0 and 2.4 regarding $F_{\rm LTW}$, depending on the model, whereas
at 1500~km, that factor varied from 1.6 to 2.3.  Different initial wave energy fluxes
were most consequential in the low and mid chromosphere, with the difference starting
to fade with increasing atmospheric height, as expected based on the limiting shock
strength behavior \citep{cun04}.  Spectral waves entail a somewhat larger wave energy
flux at large heights as those models exhibit a smaller spatial decrease of the
atmospheric density.

The formation heights for Ca II and Mg II range were found to be between 700~km and
1800~km, depending on the model.  Radiative energy losses were identified to be most
pronounced behind strong shocks owing to the impact of shock-shock interaction and in
models with time-dependent hydrogen ionization omitted; i.e., models based on
non-fully time-dependent ionization (NTDI).
Shock-shock interaction entails the merging of shocks, which leads to the formation
of shocks of large strengths.  They result in strong heating events as well as significant
large-scale cooling due to momentum transfer \citep[e.g.,][]{cun87,car92}.

Peaks of Ca~II and Mg~II emission behind shocks (including those of large strengths) are
absent in TDI models owing to the difference in the time-scales between the ionization
processes and shock propagation.   Additionally, the mean atmospheric temperatures
were found to be highest in monochromatic wave models compared to spectral wave models;
this is a consequence of quasi-adiabatic cooling owing to the momentum transfer by
strong shocks.  This result agrees well with previous work by, e.g., \cite{car92,car95,car02},
on wave simulations for the Sun.

Figure~2 depicts a snapshot series for a LTW TDI model based assuming spectral
waves with an initial energy flux of $F_{\rm LTW} = 1.7 \times 10^8$ erg~cm$^{-2}$~s$^{-1}$
(see Sect. 2.2).  Each figure panel features a dominant shock formed by
shock merging.  This process can be examined based on the time-dependent and
height-dependent development of individual shocks.  For example, shock number 37,
inserted at an elapsed time of 1500~s, shows an increase in strength $M_{\rm sh}$
given as 1.23, 4.32, and 6.15 at elapsed times of 1550~s, 1600~s, and 1650~s,
respectively.  The relative density jumps $\eta$ across the shocks are identified as
0.34, 2.44, and 2.71.

It is also found that Mg is completely ionized to Mg~II, except in parts of the
stellar photosphere.  However, the Ca~II/Ca ionization degree decreases from
100\% (mid chromosphere) to notably lower values at larger heights.  At 2000~km,
for the time-steps depicted in Fig.~2, we find values between 60\% and 95\%
due to the ionization of Ca~II to Ca~III.  However, there is no noticeable change
in Ca and Mg ionization degrees between the pre-shock and post-shock regions
owing to the non-instantaneous nature of ionization; see, e.g., \cite{kne80}
and subsequent studies for detailed information.

%++++++++++++++++++++++++++++++++++++++++++++++++++++++++++++++++++++++++++++++++

\subsection{Comparisons with Observations}

A key aspect of this study is the comparison between the results of our theoretical
models, based on both ACWs and LTWs, and observations.  Table~3 summarizes the
results obtained for the Ca~II H+K fluxes from the various models and the observations
(see Appendix~A for details).  The prime observational value is
given as log~$F_{\rm Ca~II~H+K}$ = 5.835, corresponding to the approximate
chromospheric minimum value without the photospheric contribution.  Although
our theoretical models only provide the Ca~II~K fluxes, we use the conversion
$R_{\rm K} / R_{\rm H} = 1.04$ amounting to $R_{\rm K} + R_{\rm H} / R_{\rm K} = 1.96$.
This formula utilizes previous theoretical work by \cite{cun99} and references therein
indicating the $R_{\rm K} / R_{\rm H}$ is highest for stars of largest chromospheric
activity, corresponding to fasted rotation, i.e., lowest value of $P_{\rm rot}$.
Table~3 conveys that LTW models based on $f_0$ = 0.3\% appear to be most consistent
with the observed Ca~II H+K fluxes.

%  Figure 4
%%%%%%%%%%%%%%%%%%%%%%%%%%%%%%%%%%%%%%%%%%%
%
\begin{figure*}
\centering
  \includegraphics[width=0.60\linewidth]{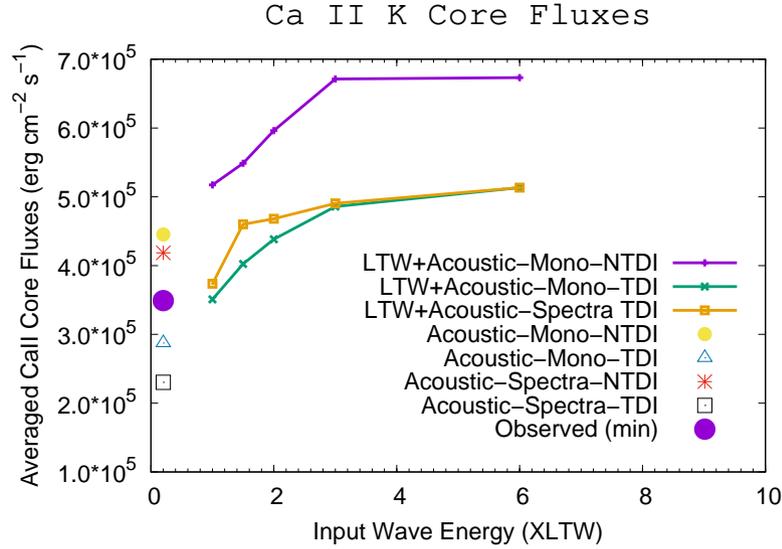}
  \caption{Averaged Ca~II~K core fluxes for different types of models as a function of
multiples of LTW wave energies based on $f_0$ = 0.3\%.
Results are given for monochromatic NTDI and TDI models,
as well as spectral TDI models.  Moreover, we also give the results for the various models
based on acoustic waves as well as the result from observations (approximate minimum value).}
\end{figure*}
%%%%%%%%%%%%%%%%%%%%%%%%%%%%%%%%%%%%%%%%%%%

%  Table 3
%%%%%%%%%%%%%%%%%%%%%%%%%%%%%%%%%%%%%%%%%%%
%
\begin{table}
	\centering
	\caption{Comparison of Ca~II~H+K Fluxes}
\begin{tabular}{lclc}
\noalign{\smallskip}
\hline
\noalign{\smallskip}
Model  &  MFF  &  Description  &  Result   \\
\noalign{\smallskip}
\hline
\noalign{\smallskip}
Acoustic                &  ...      & Mono-TDI      &  5.75   \\
...                     &  ...      & Spectra-TDI   &  5.65   \\
1.0~LTW                 &  $0.3\%$  & Mono-TDI      &  5.84   \\
...                     &  $0.3\%$  & Spectra-TDI   &  5.86   \\
1.0~LTW                 &  $1.4\%$  & Mono-TDI      &  6.05   \\
...                     &  $1.4\%$  & Spectra-TDI   &  6.11   \\
2.0~LTW                 &  $0.3\%$  & Mono-TDI      &  5.93   \\
...                     &  $0.3\%$  & Spectra-TDI   &  5.96   \\
3.0~LTW                 &  $0.3\%$  & Mono-TDI      &  5.98   \\
...                     &  $0.3\%$  & Spectra-TDI   &  5.98   \\
Observed (min, approx.) &  ...      & ...           &  5.835  \\
\noalign{\smallskip}
\hline
\noalign{\smallskip}
\multicolumn{4}{p{0.95\columnwidth}}{
Note:
The results are given in logarithmic units.
}
\end{tabular}
\end{table}
%%%%%%%%%%%%%%%%%%%%%%%%%%%%%%%%%%%%%%%%%%%

Figure~4 displays the Ca~II~K core fluxes
(defined over a wavelength range of $\Delta\lambda = \pm 0.3$~{\AA})
for different types of models together with
the observed value.  It shows again that the two-component chromosphere model
(ACW + LTW) with $f_0 = 0.3${\%} is most consistent with the observations.
Models based on pure acoustic waves yield Ca~II~K core fluxes that are notably
too small.  This figure also shows that Ca~II~K core fluxes based on models with
TTWs (see Paper~I for details), according to the approximation as used here,
are considerably too high in order to successfully describe the observed minimal
Ca~II chromospheric emission.  The models exhibit a notably increased mechanical
energy flux in the Ca~II line formation region accounting for the difference (see
Table 2).  However, these models are expected to be relevant for stages of
higher activity as previously also obtained for 55~Cnc.

%++++++++++++++++++++++++++++++++++++++++++++++++++++++++++++++++++++++++++++++++
%++++++++++++++++++++++++++++++++++++++++++++++++++++++++++++++++++++++++++++++++

\section{Summary and Conclusions}

We pursued various chromospheric heating calculations for 55~Cnc,
a main-sequence star of spectral-type late-G taking into account that
this star generally exhibits a low level of chromospheric activity.
According to previous work by, e.g., \cite{lin83} and others,
magnetic heating in stellar atmospheres is strongly correlated
to increased outer atmospheric emission.  For 55~Cnc, the magnitude
of magnetic heating is expected to be relatively low.  This behavior
stems from its old age \citep{mam08,bra11,yee17} and slow rotation
\citep{hen00,bou18}.  Stellar evolution models indicate that
late-type main-sequence stars gradually lose angular momentum
resulting in a steady decrease of stellar activity
\citep[e.g.,][]{cha97,mit18}.

Our set of theoretical models is based on detailed studies of ACWs
and LTWs.  Both kinds of waves form shocks leading to modified
thermodynamic atmospheric structures and emergent Ca~II~H+K
and Mg~II~{\it h}+{\it k} emission line fluxes.  Using Ca~II~K as
an appropriate representation of the overall chromospheric emission flux,
the line has been calculated in detail (multi-ray treatment) assuming PRD
in combination with TDI (main models).  Furthermore, our study makes
use of previous work by \cite{mus94,mus95,ulm96}, and \cite{faw11}
that provides detailed information on the amounts of magneto-acoustic
heating based the adopted stellar parameters.  In addition, 55~Cnc's
photospheric MFF has been estimated to be relatively low; therefore,
we calculated sets of models based on $f_0$ = 0.3\% and 1.4\%.

As expected, models of different MFFs indicate that larger MFFs correspond
to higher (on average) chromospheric temperatures. Those models also show
higher Ca II and Mg II emissions as well as higher Ca and Mg ionization rates,
especially in the upper magnetically heated chromospheres (see Paper~I for
details).  Additionally, different MFFs result in different rates of dilution of the
magnetic wave energy fluxes, which are instrumental for the energetics
mirrored by both the shock dissipation and the emergent radiative emission at
different atmospheric heights.

Comparisons with observations indicate that considering 55~Cnc's
low activity stage, the emergent Ca~II flux can best be reproduced by a combination
of acoustic heating and longitudinal flux tube heating, as stipulated by a
small photospheric magnetic filling factor informed by the star's slow
rotation rate.  Based on our master models, assuming both time-dependent ionization
and spectral waves, a MFF of $f_0$ = 0.3\% appears to be more consistent with the
Ca~II observations than $f_0$ = 1.4\%.  The latter types of models feature a
notably higher wave energy flux at a broad range of chromospheric heights,
including regions where the Ca II and Mg II lines form.  However, this result should
be revisited based on future models considering alternate distributions of flux tubes
and/or the effects of 3-D geometry, among other phenomena.

Possible effects associated with 3-D geometry include radiative transfer, spectral line
formation, (magneto-)hydrodynamic features (including shocks), and small-scale
geometry; see, e.g., \cite{sch00}.  For example, \cite{han07} discussed some of
these processes for the Sun, motivated by the overarching goal to adequately describe
how energy flux generated in the solar convective zone is transported and dissipated
in the outer solar layers.  One important difference is that in 3-D models, the build-up
of strong shocks due to shock-shock interaction is considerably reduced,
which affects both the heating rate in the post-shock regions and the
large-scale cooling initiated by the shock-related momentum transfer.

Our study was aimed at exploring the heating processes associated with minimal
stellar activity.  For 55~Cnc we found that for stages of lowest chromospheric activity
the observed Ca~II fluxes are similar, but not identical to those given by acoustic
energy dissipation; however, agreement can be obtained if low levels of magnetic heating
are considered as an additional component; see, e.g., \cite{sch87} and \cite{rut91}
for early results on underlying heating processes.  This outcome is consistent with
the paradigm that even stars of slow rotation are expected to have some magnetic field
coverage remaining, which is able to add to the overall heating picture; see, e.g.,
\cite{sch18}, and related studies.

%++++++++++++++++++++++++++++++++++++++++++++++++++++++++++++++++++++++++++++++++
%++++++++++++++++++++++++++++++++++++++++++++++++++++++++++++++++++++++++++++++++

\section*{Acknowledgements}

The authors acknowledge the comments by an anonymous referee who draw
our attention to additional literature on the topic of study.

\section*{Data availability}

The data underlying this article will be shared on reasonable request to the corresponding author.

%%%%%%%%%%%%%%%%%%%%%%%%%%%%%%%%%%%%%%%%%%%%%%%%%%



%\bibliographystyle{mnras}

%%%%%%%%%%%%%%%%%%%%%%%%%%%%%%%%%%%%%%%%%%%%%%%%%%

%++++++++++++++++++++++++++++++++++++++++++++++++++++++++++++++++++++++++++++++++

\appendix

\section{Stellar Basal Flux}

Following \cite{mid82} and \cite{rut84} --- see also \cite{cun99} and \cite{mit13} for
background information --- the total (physical) surface flux in the cores of the Ca~II H+K lines
$F_{\rm HK}$, encompassing both chromospheric and photospheric contributions, can be related to
the Mount Wilson Observatory S-index, also denoted as $S_{\rm MWO}$, by
\begin{equation}
F_{\rm HK} \ = \ C_{\rm cf} \cdot T_{\rm eff}^4 \cdot 10^{-14} \cdot K \cdot S_{\rm MWO}
\end{equation}
where, as for cool main-sequence stars,
\begin{equation}
\log{C_{\rm cf}} \ = \ 0.24 + 0.43 (B-V) - 1.33 (B-V)^2 + 0.25 (B-V)^3
\end{equation}
is a color-dependent, empirical transformation term
and in this historical context $K$ has been supposed to be a constant.

However, \cite{mit13} showed by using self-consistent physical atmospheric models
based on the PHOENIX code, see \cite{hau99} for further information, that $K$
is not a constant but a somewhat color-dependent term as well, given the original
$C_{\rm cf}$ as conveyed above.  As part of this work, surface fluxes in reference
windows of the S-index definition have been quantified.
For cool main-sequence stars the equation
\begin{equation}
\log{K} \ = \ 6.086 - 0.2088 (B-V) + 0.3564 (B-V)^2 - 0.002 (B-V)^3
\end{equation}
reconciles the above-given empirical narrow band photometric calibration by \cite{noy84}
with up-to-date atmospheric models, including the stellar surface fluxes.

For 55~Cnc, with $(B-V)=0.87$ and $T_{\rm eff} = 5165$~K, we find $C_{\rm cf}=0.592$
and $K=1.49 \times 10^6$. The basal flux value identified by \cite{bou18} is given as
$S_{\rm MWO, basal} = 0.14$, consistent with the previous value by \cite{bal97}.
This value is furthermore consistent with the solar basal S-value
of 0.15 and the empirical decline of $S_{\rm MWO, basal}$ towards
cooler main sequence stars, as shown by \cite{sch12} based on the lower envelope of the
\cite{dun91} S-index distribution.  Hence, this is equivalent to a total line core
surface flux during a minimal ``basal'' chromospheric emission level of
\begin{equation}
F_{\rm HK}^{\rm basal} \ = \ 8.77 \times {10^5}~{\rm erg}~{\rm cm}^{-2}~{\rm s}^{-1} \ .
\end{equation}

Even today it is difficult to physically quantify the photospheric line core flux in a meaningful manner.
Most atmospheric models do not take into account the chromosphere, thus defining the temperature minimum
somewhat arbitrarily.  In addition, UV surface fluxes of cool stellar atmospheres
also depend on the correct knowledge of the effective temperature; see, e.g., \cite{lin80} and subsequent
work for details.  Thus, the computed Ca~II H+K line photospheric core fluxes may not exactly match the
observed spectrum around and in those deep line cores.

For example, an empirical of the photospheric Ca~II line core flux has been developed by \cite{noy84}
and \cite{har84}; see also \cite{mit13} for further information.  This latter approach given as
\begin{equation}
\log{{\rm F_{HK}^{phot}}} \ = \ 6.19 - 1.04 (B-V)
\end{equation}
appears to be straightforward and consistent.  For 55~Cnc this result reads
\begin{equation}
F_{\rm HK}^{\rm phot} \ = \ 1.93 \times {10^5}~{\rm erg}~{\rm cm}^{-2}~{\rm s}^{-1} \ ,
\end{equation}
amounting to a purely chromospheric basal flux (i.e., with the photospheric contribution being removed) of
\begin{equation}
F_{\rm HK}^{\rm chrom, basal} \ = \ F_{\rm HK}^{\rm basal} - F_{\rm HK}^{\rm phot, basal} \ = \
6.84 \times {10^5}~{\rm erg}~{\rm cm}^{-2}~{\rm s}^{-1} \ .
\end{equation}
This value is consistent with the prediction based on the Rossby number as given by, e.g., \cite{noy84} and
\cite{mon01}.  Note that compared to the Sun the photospheric core flux decreases more rapidly toward
main-sequence stars of lower effective temperature such as 55~Cnc, relative to the chromospheric basal flux.
Consequently, the latter appears to be more pronounced in the Ca~II H+K line cores for those stars.

%++++++++++++++++++++++++++++++++++++++++++++++++++++++++++++++++++++++++++++++++

% Don't change these lines
\bsp	% typesetting comment
\label{lastpage}
\end{document}